\documentstyle[psfig]{mn}
\begin{document}
%
\title[Warp in the $\beta\:$Pictoris~disk]
{A planet on an inclined orbit as an explanation of the warp in the $\beta\:$Pictoris~disk}
\author[]
{ D. Mouillet$^{1}$ \and
 J.D. Larwood$^{2}$ \and J.C.B. Papaloizou$^{2}$ \and A.M. Lagrange$^{1}$\\
$1$ Laboratoire d'Astrophysique de l'Observatoire de Grenoble, UMR 5571,
Universit\'e J. Fourier, BP 53,
F-38041 Grenoble Cedex 9\\
 $2$Astronomy Unit, School of Mathematical Sciences, Queen Mary \& Westfield
College, Mile End Road,\\
 London E1 4NS}

\date{Received: ; accepted: }

\maketitle

\def\km{km\,s$^{-1}$ }
\def\h{\hfill\break}
\def\bp{$\beta\:$Pictoris}
\def\cs{circumstellar}
\def\mic{$\mu $m}
\def\arcsec{\hbox{$^{\prime\prime}$}}
\newcommand \cop{\mbox{\sc come-on-plus}}
\newcommand \ad{\mbox{\sc adonis}}

\begin{abstract} We consider the deformation that has recently been
observed in the inner part of the circumstellar~disk around
$\beta\:$Pictoris~with the HST.  Our recent ground based adaptive optics
coronographic observations confirm that the inner disk is warped.  
We investigate the
hypothesis that a yet undetected planet is responsible for the observed
warp, through simulations of the effect of the gravitational perturbation
due to a massive companion on the disk. The physical processes assumed in
the simulations are discussed: since the observed particles do not survive
collisions, the apparent disk shape is driven by the underlying
collisionless parent population.
The resulting possible parameters for the
planet that are consistent with the observed disk deformation are reviewed.
\end{abstract} 

\section{Introduction}
An important discovery by IRAS was the detection of IR excess due to cold
material around a number of Main Sequence Stars. Among them, \bp ~exhibits
one of the largest excess (Aumann, 1985).  Shortly after the IRAS results,
a disk of dust was imaged around this star (Smith and Terrile, 1984). Since
then, the disk has been extensively studied , as it is expected to be
related to planetary systems, possibly in a state of evolution different
from that of our Solar System.

The disk composition is complex.
Micron sized grains are
detected through scattered light (Kalas and Jewitt, 1995  and
ref. therein; Mouillet et al, 1996) and thermal emission (Lagage and Panin,
1994). 
The presence of  small amounts of submicron grains is  inferred
from 10 \mic ~spectrophotometry (Knacke et al, 1993).
Larger grains (mm sized) are detected through photometry at mm wavelengths
(Chini et al, 1991; Zuckerman and Becklin; 1993).
Kilometer-sized bodies have also been proposed to account for the very
peculiar spectroscopic variability of \bp ~(Lagrange et al, 1987).
Gas is detected spectroscopically through the presence of absorption lines. 
However the gas to dust ratio is probably $\leq 1.$, ie much less than
that in the environment of young stars such as T Tauri,
or in star forming regions. Most of the gas is probably confined very 
close to the disk (Lagrange, 1995 and ref. therein). 

The total mass of the disk is very uncertain as most of it comes
from the 
largest bodies. About one Earth Mass is necessary to account for optical to
mm observations. If the  particle size distribution follows a  law of the form,  
$$dn(a) \propto  a^{-3.5}da,$$ 
with $n(a)$ being the number of particles of size greater than $a,$ 
up to kilometer size bodies, the mass would be several tens of Earth Masses
(Backman and Paresce, 1993).  
 
The origin and evolution of this system is still the subject of
active research and one point of investigation is the possible presence of
planets within the disk.
From the theoretical point of view, planets are
expected to be formed out of the the circumstellar disk that accompanies
the star during the process of formation. The timescale estimated
for this process (see Lin and  Papaloizou, 1985)
is characteristically less than
the estimated lifetime of
 \bp ~$\sim 2.10^8 y$ (Paresce, 1991). 
Thus is is reasonable to suppose that planets
have had time enough to form.

 Obtaining  direct evidence for the existence of  planets around early type stars such as \bp
~ is far from straightforward. Direct imaging is still  beyond 
 current observational capability because of the high contrast between
the star and the planet. Photometric variations have been observed
(Lecavelier et al, 1995) which are consistent with the occultation of \bp
~by an orbiting Jupiter-like planet. But this has not been firmly established
(observation of another occultation is needed to sweep all
doubts away). 
Radial velocity studies, such as those performed to detect giant planets
around solar type stars such as 51 Peg, 47 UMa, and others (Mayor and Queloz, 
1995, Marcy and Butler, 1996) would not provide 
enough accuracy to detect giant planets at distances of about a few AU from these
ususally rapidly rotating early type stars.

Signatures of planets  can result from  their  gravitational effect on the
disk. The observed
clearing of the inner region of the disk
has been attributed to the gravitational
effect of an orbiting planet (Roques et al, 1994).  
But other physical processes can 
explain the observed dust distribution as well.

The gravitational  influence  of a planet has been invoked in order to
explain the observed high rate of cometary infall (about 1000 per year
corresponding to $\sim 10^{-15} M_{\odot} y^{-1}$). Beust and Morbidelli
(1995) propose that mean motion resonances between the orbits of kilometer
sized bodies and a planet in a longer period eccentric orbit  can  produce
the right quantity,  and orbits of infalling bodies. They predict long term
variabilities which are currently  being tested via a spectroscopic survey.
Such a model requires a steady flux of kilometer sized bodies from the
outer to inner regions of the disk to occur as a consequence of collisions.

Finally, HST data provide high S/N data on the light scattered
from the the inner part of the disk down to 25 AU from the star
(Burrows et al, 1995). They revealed  a slight inclination ($\sim$ 3
degrees) of 
the inner disk midplane up to about 50 AU,  to the outer disk
midplane. Burrows et al (1995) proposed that a planet on an  orbit inclined to
the outer disk plane might be  the most probable explanation for this
observation. In section 2, we   show  recent ground-based
 observations imaging the same region of the disk for comparison. 
In section 3 we review the physical properties of the disk to determine how
an hypothetical planet would act on it. 
This leads to 3-D numerical simulations  of
the simultaneous behaviour of large numbers of test particles representing
a disk population of kilometer sized objects under the gravitational field
of the star and  massive companion on an inclined, and possibly eccentric
orbit. In section  4 we present the results. We finally discuss the
possible parameters for the companion 
  consistent with its presence being able
to explain the brightness asymmetry of the inner \bp
~disk.

\section{Observed asymmetry of the inner part of the \bp ~disk}
\subsection{Adaptive optics observations of the inner disk }
No warp has been detected in the classical coronographic observations
outside 80 AU. In order to observe the disk closer to the star, one needs
high angular resolution observations, ie  either space observations or
atmospheric effect corrected images. Recently, adaptive optics
coronographic images of the \bp ~disk enabled us to detect the disk in the
near infrared (NIR) through
scattered light down to about 25 AU from the star  (Mouillet et al,
1996). Similar resolution  was obtained with HST in the optical range
(Burrows et al, 1995). 

We performed new observations of the \bp ~disk with a coronograph coupled
 to the ESO adaptive optics system ADONIS on January 5, 1996 in the J
 band (Fig. 1).
\begin{figure*}
\psfig{file=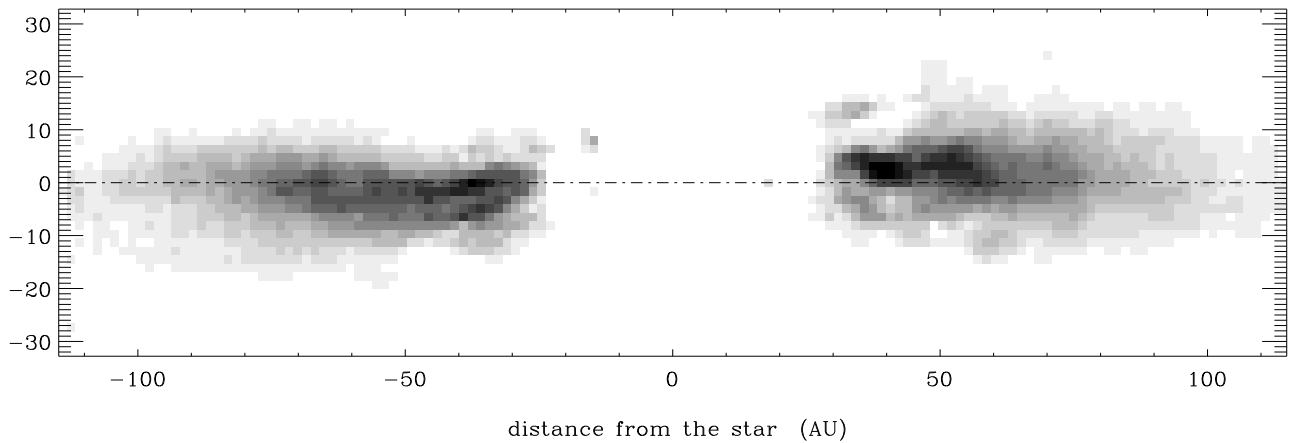,width=17cm,angle=90}
\caption{Adaptive optics observations in the J band of the inner disk of
  \bp , January 05 1996. The axes are marked in AU. Inward of  50 AU, the disk midplane
  is inclined  with respect to the outer disk midplane (dashed line)}
\end{figure*}
The inner region (30-80 AU) is observed with an angular resolution similar to
that of HST. However, the correction for atmospheric turbulence makes it
more difficult to subtract perfectly the stellar light which remains after
the mask,  so that some residuals of the PSF temporal variations remain on
the image. Details of the observing method and of the reduction procedure
can be found in Beuzit et al (1996).

\begin{figure}
\psfig{file=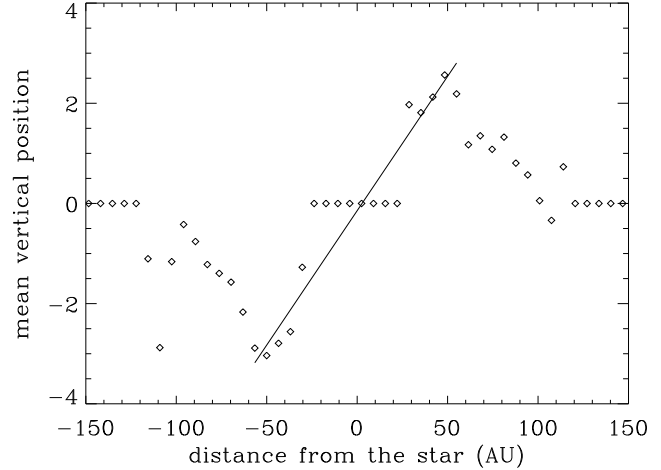,height=6.2cm,angle=90}
\caption{The observed vertical deformation of the disk above  the outer
  disk midplane. The axes are marked in AU. The vertical deformation is
  measured  from the centroid of the brightness distribution in the
  direction normal to the outer  disk midplane. The  value  zero is adopted
  when the measurement 
  is not possible (closer than 25 AU and further than 110 AU). The solid
  line is the linear fit  for the inner part. We define
  the extent of the deformation as the distance  at which the slope of
  the fit starts  to decrease. The extent of the deformation is  50 AU,
  and the corresponding inclination 3 degrees}
\end{figure}
 From Fig.2 we see that   the intensity peak  lies on one of  two  
lines according to whether it is  nearer  or further
than 50 AU from the centre. Thus the disk midplane  inside 50 AU is inclined at about 3
degrees to the outer disk midplane. The effect of the  warp  is  hardly
seen, but nonetheless it is consistent with the HST data in terms of
amplitude and extent.

These observations confirm previous results (Mouillet et al, 1996; Burrows
et al, 1996): the surface brightness profile $I(r),$  which is measured to
be very 
steep far from the star (Kalas and Jewitt, 1995), gets flatter and flatter
as the distance to the star, $r, $  decreases from 100 AU down to 25 AU.
 For $r >$ 100 AU,  I(r)$\propto r^{-3.5}$. For  50 AU $ <r< $ 100 AU,
 I(r)$\propto r^{-1.3}.$ For $ r <$ 50 AU, the profile is  even flatter.
(Fig. 3). 
\begin{figure}
\psfig{file=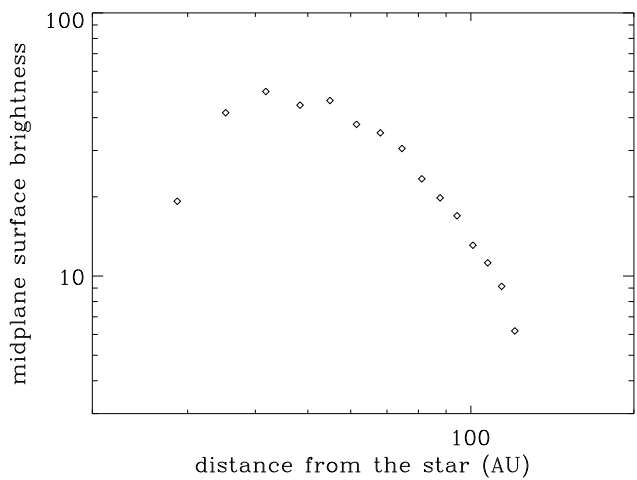,height=6.2cm,angle=90}
\caption{Midplane surface brightness of the disk}
\end{figure} 
Consequently, the form of the optical optical  depth   as a function of 
 $r,$ given by
 $\tau (r) = 5\,10^{-3} (r/100)^{-1.7},$
which is valid for $r >$ 100 AU,  cannot be used for smaller values of $r.$ 
The inferred  form depends on the precise
model  involving a combination of thermal IR and scattering
observations,  but it  converges on a progressive flattening up to a maximum
value of 10$^{-3}$-10$^{-2}$ at  around $r=$  40 AU, thereafter decreasing  inwards. Very
little residual dust is expected closer than 15 AU, since it would induce
IR excesses in NIR which are not observed.
\subsection{The planetary hypothesis}
The deformation of the inner disk midplane relative to the outer disk midplane is not the
only asymmetry observed in the \bp ~disk (Kalas and Jewitt
1995; Lagage and Pantin, 1994). All these asymmetries are 
difficult to explain in the context of a low mass disk consisting of many
small objects under a dominant potential arising from a central point mass,
since any azimuthal inhomogeneities would be expected to disappear on an
orbital timescale  (a few 1000 years at 100 AU). A single physical process
is unlikely to   be responsible for all of them since very different spatial scales are
involved together with different geometries. For instance, anisotropic
scattering properties together with a slightly inclined disk were  proposed
to explain the so called ``butterfly'' asymmetry (Kalas and Jewitt, 1995:
the midplane is not an axis of symmetry of the brightness
distribution). But such an explanation is unable to count for any
radial-dependent asymmetry. 
Another idea is that  the local
production of very small grains as collision products followed by  their 
efficient accelaration by radiation pressure could produce catastrophic
blow out in some random part of the disk (Artymowicz, 1996). Such a
mechanism could explain any kind of asymmetry without the need of a
planet. But the strong antisymmetry ( global azimuthal structure)
 of the observed warping  with respect to
the central star make this an implausible explanation for it. 
In this paper we investigate the possibility that the warping asymmetry
is produced by an as yet undetected planet in order to see what
constraints might be put on its parameters. Although we are not able to claim
this as a unique explanation,  we are able to show that the required parameters
are reasonable and  not at present in conflict with constraints 
arising from the lack of observational detection through either
imaging or spectroscopy.
\section{Numerical simulations  of the influence of  a
  planet}
\subsection{Physical processes in the \bp ~disk}
The dynamics in the disk is dominated by the  gravitational force
of the central star  to which may be added   effects due to any  possible
perturbing  planets. The mass of observed disk material is negligible.
Radiation pressure  is important for small grains less than  a
characteristic size a$_c =$ 2\mic ~(Artymowicz, 1988). For such grains the
force due to radiation pressure exceeds gravity so that they are on
unbound hyperbolic orbits. Accordingly their lifetime is on the order of
the crossing time of the disk and thus very short.  The Poynting Robertson
effect is  negligible as is gas drag due to the small quantity of gas
present (Artymowicz, 1995).  

The main quantity determining the grain  collision frequency is the optical
depth of the dust, $\tau$. The mean time between collisions is given by
$$t_c ={P_{ORB}\over \pi \tau},$$
where $P_{ORB}$ is the local orbital period.
For the \bp ~disk, the optical depth has been estimated  to be typically
$5\times 10^{-3}$ for small size scattering particles. 
Supposing that the  particle size distribution is given by
$dn(a)  \propto a^{-3.5}da,$ ( which might arise from collision processes) the
smallest micron sized grains contribute most of the optical depth.
The collision time at $100 AU$ is then  typically $10^5 y$ for such grains.
The characteristic impact velocities are $\sim v_r = i v_{orb} \sim 1 km/s
,$ where $v_{orb}$ is the orbital velocity and  $i$ is a characteristic
inclination in radians associated with a grain orbit.
The expected impact velocities are sufficient to destroy the  colliding
particles.  Therefore the lifetime of the  smallest  particles in bound
orbits 
is on the order of the collision time $\sim 10^5 y$ at 100 AU 
and even shorter closer
than 100 AU, in any case  much less than the age of the system.
Thus they need to be replenished by the destruction of larger
particles. For normal solid state densities, the surface density of disk
material contained in small particles is $\Sigma \sim a_c \tau,$ which implies
a mass of $\sim 10^{-9} M_{\odot}$ contained within $100 AU,$ and a mass
loss rate of around $10^{-14} M_{\odot} y^{-1}.$ 

Assuming
 that  collisions will  determine both the particle size  distribution through
fragmentation processes $(dn(a)  \propto a^{-3.5}da)$  as well as the particle
lifetimes,  most of the mass in the
system will reside in the largest bodies which also contribute least to the
optical depth. 
It is natural to regard the larger bodies, as a primary source of
material to drive the global mass loss rate of the system. After a
(destructive) collision, the  particle ejection velocities, as viewed
in the centre of mass frame,  are negligible
compared to the orbital velocity of the parent particles. 
Consequently, the global distribution of any given size particles is not
driven by interaction processes but is closely related 
to that of the much bigger parent bodies. 
This means that the disk shape of the small particles detected in scattered
light is driven by the kinematic effect of gravitation on long-lived
(old as the system) particles .


Because of the detection of this disk, the collision time of the biggest
bodies is expected to be larger than the age of the system. 
As an illustrative example, if the largest sized bodies are
$km$ sized, the given size distribution indicates that $\sim 3.10^{-5}
M_{\odot}$ should  be contained in these bodies and the characteristic
collision 
time between them to be $\sim 3.10^9 y.$ If a significant amount of the
material involved in collisions ends up in  small particles, then the
supply rate could be as much as $ \sim 10^{-14} M_{\odot} y^{-1}.$ Note too
that collisions between the $km$ sized bodies also result in an inward
radial migration of the distribution  much as for viscous accretion disks
or planetary rings. The effective diffusion coefficient is $ D_c \sim H^2
/t_c,$ 
$H$ being the vertical thickness and $t_c$ being the collision time. For
$H=0.1r,$ and $t_c = 10^9 y,$ this gives a diffusion timescale of $r^2/D_c
\sim 10^{11}y,$ corresponding to a mass flow rate of  $\sim 3.10^{-16}
M_{\odot}y^{-1}.$ 
Apart from lifetime considerations, a large number of such $km$ sized
bodies is expected in the \bp ~disk to explain the spectroscopic detection
of highly redshifted variable gas close to the star (Beust et al. 1996). 
Were even larger bodies to exist around the star, like
the kilometer sized objects they might also provide a source of small
particles, but they would be expected to have a similar
or perhaps even longer effective collision frequency, and thus behave in a
similar way 
from the point of view of our analysis.

Accordingly we model the disk as a system containing a large number of
long-lived objects, which as far as the majority is concerned, are
collisionless.  We then suppose that, as they do not survive more than one
collision,the small particles reflect the distribution of the large
bodies. We are thus concerned with kinematic patterns produced by
non-interacting particles. 
\subsection{Numerical simulations}
%
The numerical simulations are based on a collisionless adaptation of 
the SPH code described in Larwood and  Papaloizou (1997, and references
therein). In this  
version pressure and viscous forces are removed so that we consider  a
purely kinematic model. 
The code follows the velocity and position coordinates of typically 15,000
particles over thousands of revolutions of the perturbing companion. The
particles were initially set up in circular motion in a disk configuration with
aspect ratio $H/r = 0.1.$ They were inserted, using a random number generator
 according to the following
$\Sigma$ distribution (in model units) to schematically represent the
matter distribution in the \bp ~disk (since the particles are
collisionless, the precise form of $\Sigma$(r) has no effect on the
amplitude and extension of the consequent warp):\h 
- for 1.2 $\leq$ r $\leq$ 4.2, $\Sigma$(r) $\propto r^{-0.5}$\h
- for 4.2 $\leq$ r $\leq$ 6, $\Sigma$(r) $\propto r^{-1.5}$.\h
The model  unit of length corresponding to $r=1,$ $R_{unit},$ is arbitrary, whereas the corresponding
time unit $t_{unit}$ is the inverse orbital frequency at $R_{unit}$.

The companion is  initiated in a generally eccentric orbit such
that, in a Cartesian coordinate system  with origin at the center of mass,
and $(x,y)$ plane coinciding with the  initial disk midplane,
both the planet and the line of nodes are on the $x$  axis at time $t=0.$
Parameters associated with the simulations are, the mass ratio  of the
planet and central star,   $M/M_{\star},$ the semimajor axis, $D$,
eccentricity, $e$, and  inclination of  the planet's orbit, $i,$ to the
disk midplane. Finally the run time since initiation with the orbital
configuration as described above is also an important parameter as this
determines the extent of the warped deformation. Table 1 gives the chosen
parameters for various simulations. 
\begin{table}
\caption{Parameter sets of the simulations in model units (see text)}
\begin{tabular}{|l|r|r|r|r|l|}
\hline
Model&$M/M_{\star}$&D&e&i&Run time\\
\hline
01&1e-3&1.&0.&3&  1e5\\
02&4e-3&0.5&0.&3& 1e5\\
03&9e-3&0.33&0.&3&  3e5\\
04&5e-3&1.&0.&3&  1e5\\
05&1e-2&0.7&0.&3& 1e5\\
06&1e-2&1.&0.&3&  1e5\\
07&1e-3&0.7&0.1&3&
 7.5e5\\
08&1e-3&0.7&0.3&3& 3e5\\
09&1e-3&0.7&0.5&3& 3e5\\
10&1e-2&0.7&0.1&3& 
 3e5\\
11&1e-2&0.7&0.3&3& 3e5\\
12&1e-2&0.7&0.5&3& 3e5\\
12b&1e-2&0.7&0.5&6& 3e5\\
\hline
\end{tabular}
\end{table}

In order to compare the simulation results with observation, we
have derived the apparent surface brightness distribution associated with
the simulation particles. That is the integration along a line of sight of
the scattered stellar flux from each particle (Fig. 4). 
\begin{figure*}
\psfig{file=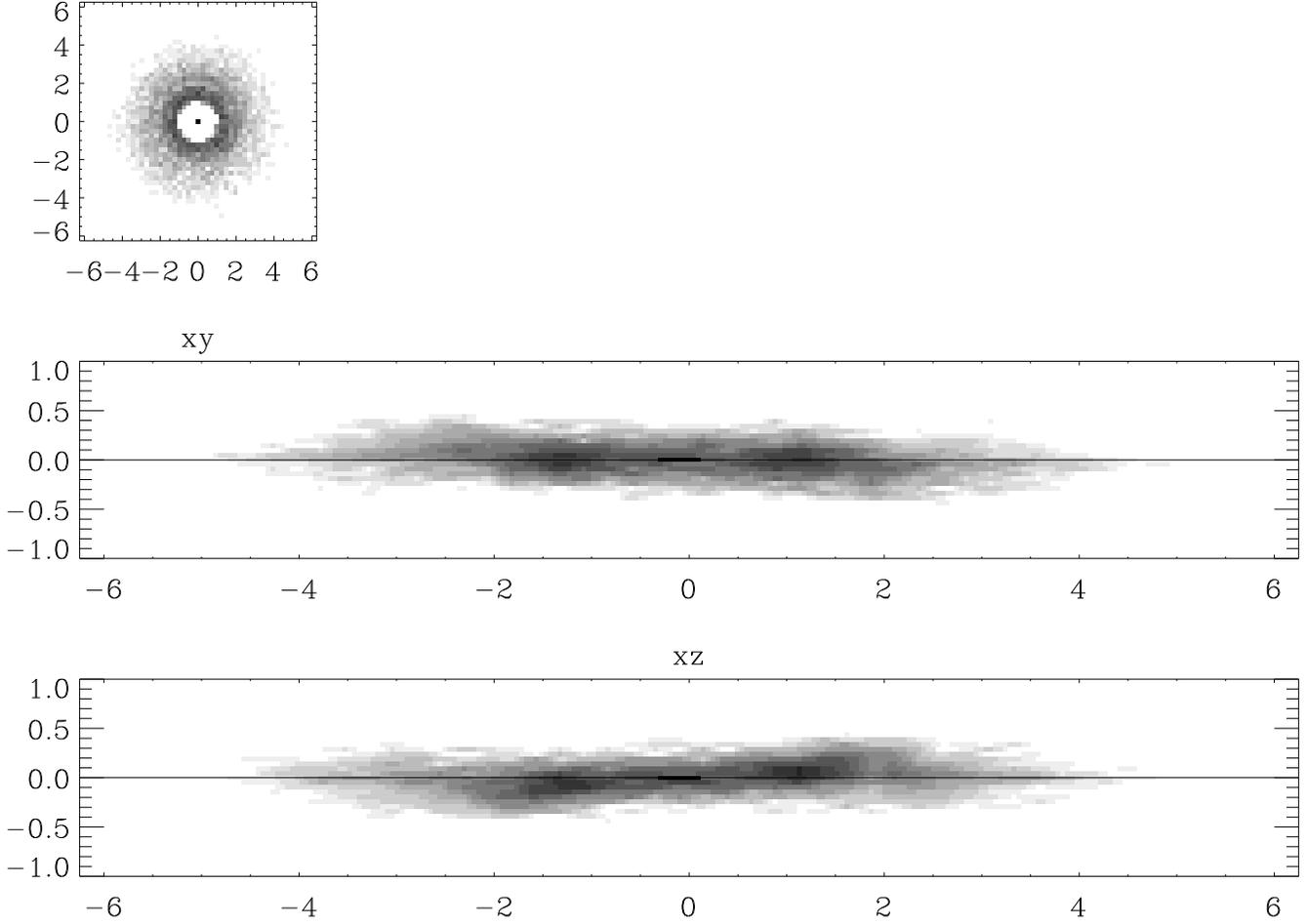,height=13cm}
\caption{Apparent surface brightness distribution associated with the
  simulation particles. The represented data set is model 05 at run time =
  10$^5$. For each projection plane, the axes are scaled in model units}
\end{figure*}
We note that a companion in
an inclined orbit  breaks the system's  azimuthal symmetry. The orientation
of the line of sight to line of nodes is  unknown in the case of \bp , and
should be considered as a free parameter in fitting the observations.
\section{Results}
\subsection{Simple description of the effect of a companion
on the vertical elevation of the disk}
Adopting the cylindrical coordinates $(r,\varphi,z),$ equivalent to our Cartesian
system,
 the vertical elevation of a razor thin disk moving under the
influence of a companion in an inclined circular orbit with small $i$ satisfies
\begin {equation} 
{d^2 z\over d t^2} +\Omega^2_z z = {3\over 4}{G M D^2\over r^4}
\sin(2i)\sin(\varphi).\label{vert}\end {equation}
Here $\Omega_z$ is the frequency of vertical oscillations and the term on the
right hand side is the secular vertical acceleration due to the companion.
This has been expanded to lowest order in $D/r,$ which becomes an increasingly
accurate procedure further out in the disk (Larwood and  Papaloizou, 1997). 
The precession frequency $\omega_p
= \Omega -\Omega_z$, is such that $ |\omega_p|
 \ll \Omega,$ where $\Omega $ is the angular velocity. For
particles in near circular orbits,
$\varphi = \Omega t + {\rm const.},$
and for small $e$ and $i$
$$\omega_p=-{3\over 4}{G M D^2\over \Omega r^5}.$$

Using the above  in $(\ref{vert}),$ we obtain the approximate solution
with zero elevation at $t=0,$

\begin {equation}
z=-{3\over 8}{G M D^2\sin(2i)\over \Omega\omega_p r^4}
\left(\sin(\varphi)- \sin(\varphi -\omega_p t)\right).
\label{vt}\end {equation}

In this purely kinematic model, the inclined orbit of the companion breaks
the midplane symmetry and forces the test particle orbits  to precess
around the companion's orbital angular momentum axis. If phase mixing due to the
$r$ dependence of $\omega_p$
is complete ($|\omega_p t|\gg 1$), this process
thickens the disk so that the aspect ratio  appears to be equal
to the orbital inclination. The vertical amplitude of the
deformation is thus directly related to the inclination of the companion's
orbit. 

This relation is  verified  numerically  for the simulation data (Fig. 5). 
\begin{figure}
\psfig{file=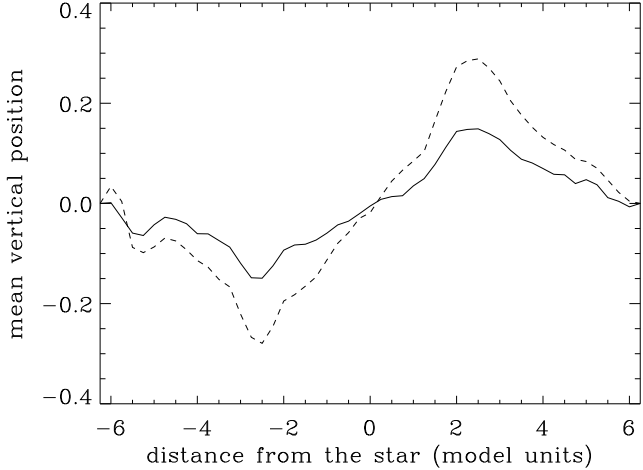,height=6.2cm,angle=90}
\caption{Linear dependence of the amplitude of the warp on the inclination
  of the companion's orbit. The axes are scaled in model units. The
  amplitude of the warp is quantified in the same way as for observational
  data (see Fig. 2). The effect is twice as large  for an inclination of 6 degrees
  (dashed line, from model 12b), compared to an inclination of 3 degrees
  (solid line, from model 12), all other parameters equal}  
\end{figure}
In addition, we note that the apparent amplitude of the warp depends on
the orientation  of the  line of sight (Fig. 6). The observability of the warp is
maximized when it (assumed to lie in the outer disk midplane)
 is at   about 30 degrees to the line of nodes.
\begin{figure}
\psfig{file=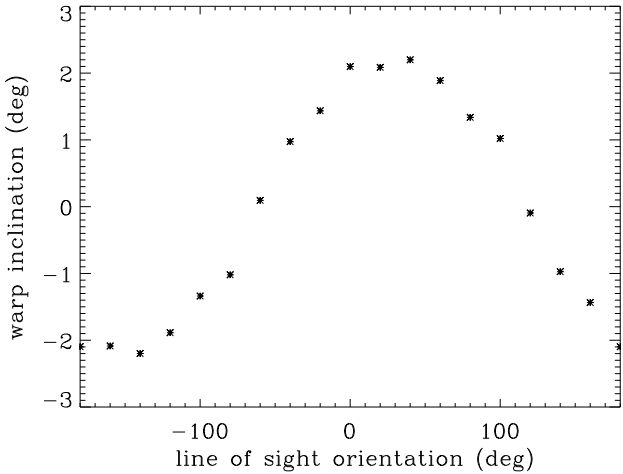,height=6.2cm,angle=90}
\caption{Observability of a warp as a function of the line of sight
  orientation. The x-axis gives the orientation of the assumed line of
  sight to line of nodes. The y-axis gives the corresponding apparent
  inclination of the warp in simulated data (with a 3 degrees inclined
  companion orbit), measured in the same way as for the
  observational data (Fig. 2).}
\end{figure}
\subsection{Radial extent of the deformation}
A deformation corresponding to a thickenned disk is established very fast close
to the star and  then propagates outwards. 
As indicated above by (\ref{vt}) the location of the end of the deformation
very roughly
corresponds to the place where the inverse precession frequency is 
roughly equal to the run time ($|\omega_p t|\sim 1$). But note that this will
overestimate the location of the edge of the deformation because many precession
periods are needed to ensure  enough phase mixing for the full deformation to be produced.
  But the number of precession periods required  should not depend on radius so that 
  we expect  the scaling
relation $$ r \propto (M D^2t)^{2/7}.$$ This form occurs because the  propagation
rate is driven by the tidal force which is proportional to $MD^2$.
 
The  radial dependence of the precession frequency, namely
$$\omega_p \propto MD^2r^{-7/2},$$
has been  checked in the simulations (Fig. 7)
\begin{figure}
\psfig{file=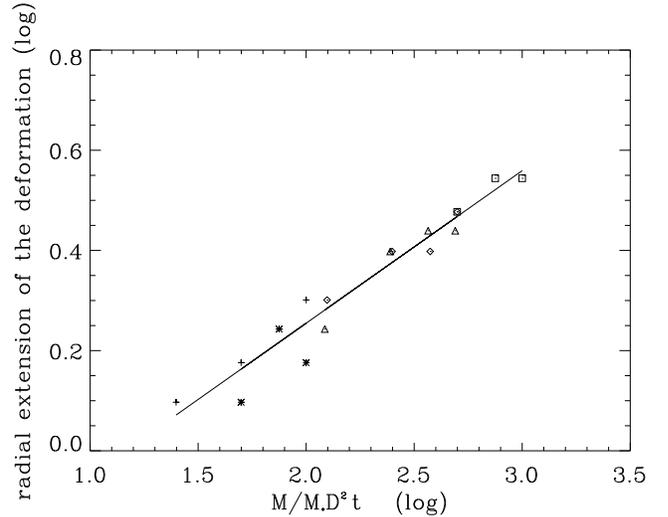,height=6.2cm,angle=90}
\caption{Radial extent of the deformation. The extent of the warp (in
  model units) is
  measured in the same way as  for the observational data (Fig. 2). It is displayed 
  versus $M D^2t$ in a
  log-log  plot. The simulation data comes from zero
  eccentricity models (model01: +, model02: $\ast$, model04: $\Diamond$,
  model05: $\triangle$, model06: $\Box$) and a line of sight orientation of 60 degrees
  (where the warp observability is unambiguous). Similar results are
  obtained with other models and orientations. The best fit power law index
  is numerically measured to be 0.29, in good agreement with the
  theoretical expectation (2/7)}
\end{figure}
\subsection{Companion eccentricity and particle inflow towards the star}
We now explore the consequence of a possible eccentricity of the
companion orbit.
Increasing eccentricities induce very little direct 
observable effect for a disk seen edge-on (Fig. 8,9).
\begin{figure*}
\psfig{file=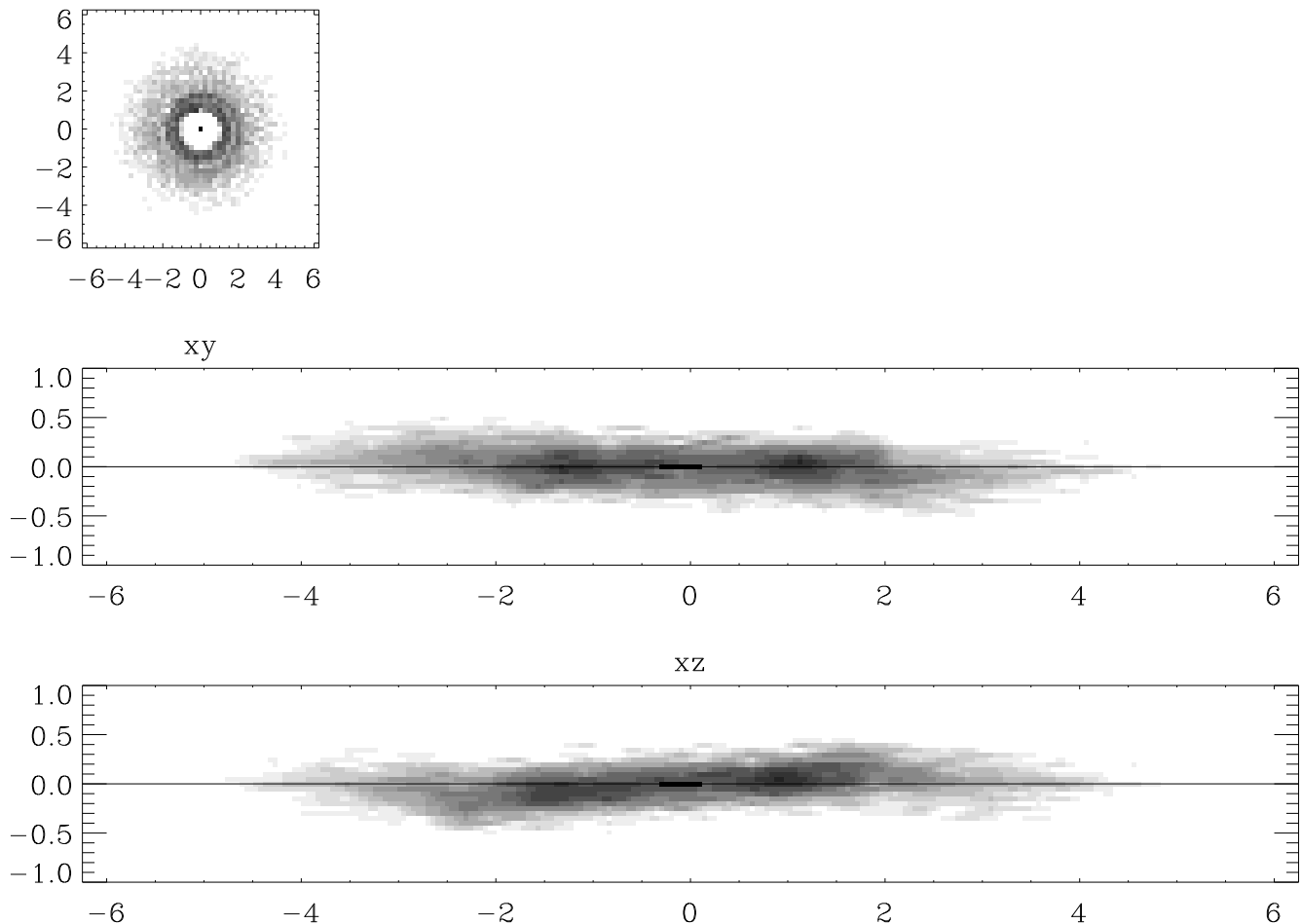,height=13cm}
\caption{Same as Fig. 4, for model10 at run time = $3\,10^5$: the companion
  orbit eccentricity is 0.1}
\end{figure*}
\begin{figure*}
\psfig{file=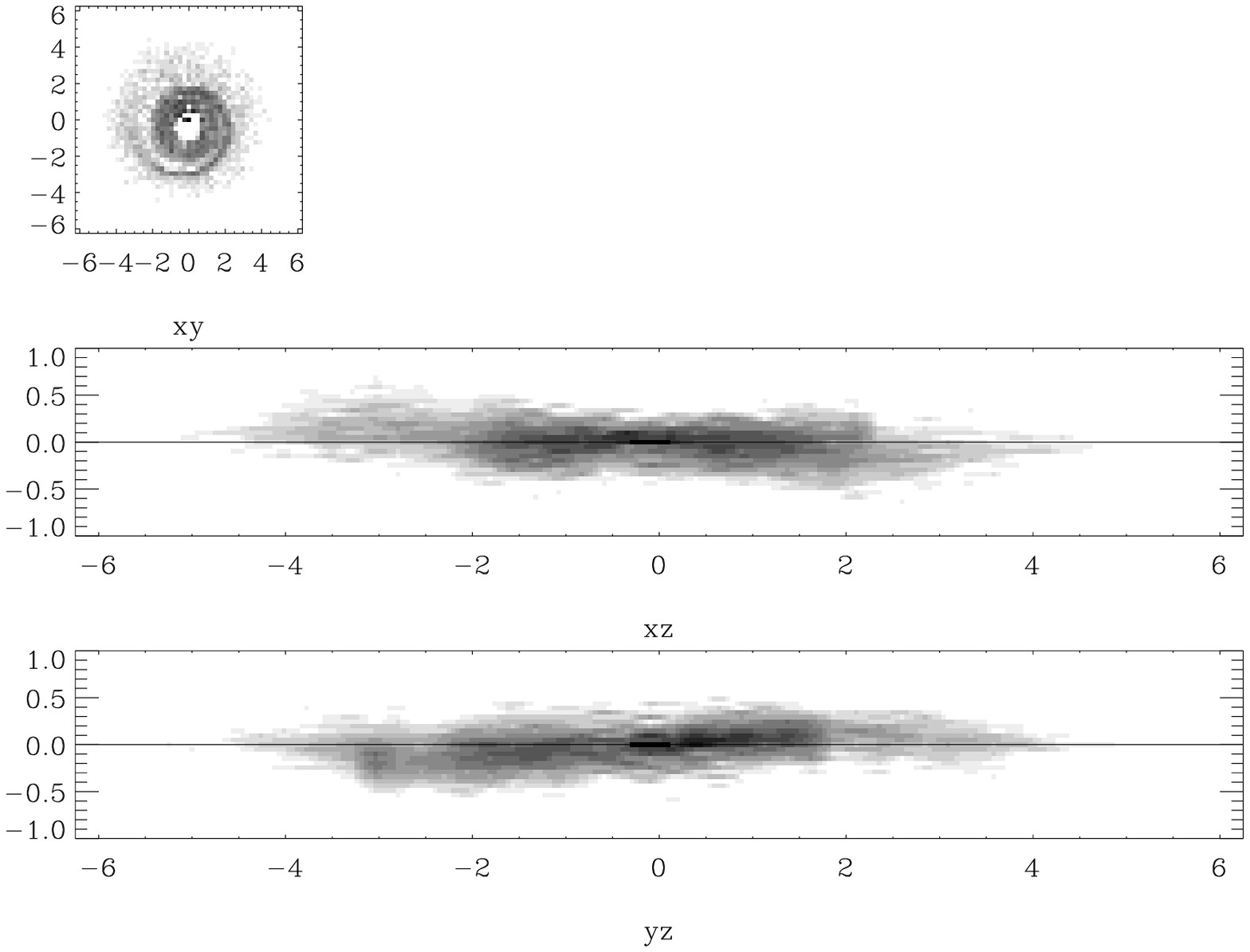,height=13cm}
\caption{Same as Fig. 4, for model12 at run time = $3\,10^5$: the companion
  orbit eccentricity is 0.5}
\end{figure*}
Consequently, this value is poorly constrained by the observations of the \bp ~disk.
Yet, the azimuthal distribution of particles is affected, with the
formation of spiral-like features. The
stronger the eccentricity, the sharper such features get.

We also notice that the eccentricity of some disk particles gets
increased to high values with the result
 that they are thrown towards the star and
end up inside the companion's orbit. Out of the initial number of 15354
particles  with $r > 1.2,$  a few  get inside the companion
orbit $(D=0.7)$ in model 11 $(e=0.3)$, and a few dozen in  model 12
$(e=0.5).$

\section{Possible parameters for the  perturber}
The comparison between the simulation results and observations allows 
derivation of   possible  parameters for the companion orbit. First, the observed
inclination of 3 degrees requires an orbit inclination of 3-5 degrees 
to the outer disk midplane, depending on the  angle between the 
line of sight  and
the line of nodes ( these assumed to be coplanar). This  angle is poorly constrained but should not
lie in the $[-80{^\circ}, -50^{\circ}]$  (modulo $180^{\circ}$) range  so that the
deformation is observable.

Second, the companion orbit is not well constrained by the observation
of an edge-on disk, so that it can be considered as a free parameter in
the simulations. This could  have an important role in feeding material into  the
inner disk. This,
taken together with the general inward migration
of the $km$ sized bodies  may be related  to 
the cometary infalls observed via spectroscopy, at the rate of about
1000 km-sized bodies per year. 
Gravitational perturbations are likely to be involved in these
processes. A massive companion invoked to explain the imaging data, could then
also provide at least part of an explanation for these infalls, as long
as its eccentricity is significantly larger than 0.1. We also remark
that a moderately large eccentricity is required if an eccentricity
pumping mechanism, through mean motion resonances,
is supposed to work on particle orbits interior to the planet.

Finally, observations show a radial extension of the warp  of 50 AU. If we arbitrarily choose
the  model unit $R_{unit}$ = 10 AU, then  the corresponding  model time
unit is :
 $$t_{unit}\; =\; \left ( \frac{r^3}{G\,M_{\star}} \right
)^{1/2} \;=\; 5.2\,y.$$
 According to the simulations (Fig. 7), the
observed extent of the warp 
(log(R/$R_{unit}$) = 0.7) requires then the condition :
 $$
\frac{M}{M_{\star}}\,\left ( \frac{D}{R_{unit}} \right )^2 \,
\frac{t}{t_{unit}} \;\sim \; 3000.$$
 Assuming that the companion formation
is rapid, so that the propagation time of the warp is  close to the
age of the system, $t \sim 2\,10^8 y$. 
 Then a planet with $M D^{2} = 10^{-4}M_{*}(10 {\rm AU})^2$  can
account for the observed warp (Fig. 10). In the framework  of these simulations,
the companion orbit is interior to the observed disk so that $D \leq 20$
AU. The absence of radial velocity variations larger than about 1 km/s in our 
spectroscopic data gathered since 1984 implies that $M^{2}/D_{{\rm AU}} \leq 
5.\,10^{-4}$. 
Meanwhile, the non direct detection of any close by
companion does not constrain further the range of possible parameters.
\begin{figure}
\psfig{file=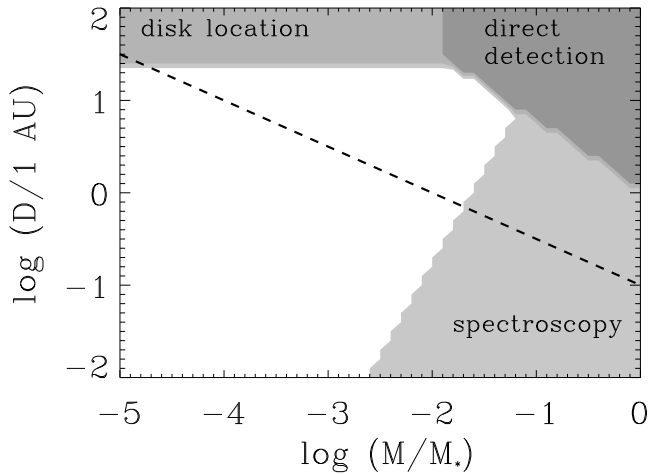,height=6.2cm}
\caption{Possible parameters M and D for the gravitational perturber in a
  log-log diagram. The constraint derived from the extension of the warp is
  represented by the dashed line. Shaded regions are forbidden because of
  observational constraints (see text) and  the 
  assumption that the companion orbit is inside the disk}
\end{figure}
Finally, the  possible parameters define a giant planet with a mass
$10^{-5} \, \leq \, M/M_{\star}  \, \leq \, 10^{-2}$, located at a 
corresponding respective  distance from the star $20 AU \, \leq \, D \, \leq \, 1 AU$.
Such a planet may be detected in photometry if it crosses the line of
sight, which requires that the earth is in the plane of the planet orbit to within  
a precision of $ 0.06^{\circ}/D_{{\rm AU}}$. Actually, Lecavelier et al
(1995) propose a Jupiter-like planet to 
explain short term light variations, which is within the present possible
parameter range. 

These quantitative results  on the possible 
companion parameters directly depend on
the system lifetime, $t$, which is still unprecisely  estimated.
However, Crifo et al (1997) derive from the stellar photometry and the new
Hipparcos data that the system may not be younger than 10$^7 y.$ In this 
case, our corresponding constraint would shift to $M D^{2} =
2\,10^{-3}M_{*}(10\, {\rm AU})^2$. 
This would restrict the possible parameters to more massive companions ($10^{-3.5} \,
\leq \, M/M_{\star}  \, \leq \, 10^{-2}$) between 20 AU and 3 AU.

\section{Conclusion}
From visible as well as  NIR imaging observations, the disk around \bp ~is
detected through scattered light  from \mic ~sized grains. These grains are
short-lived because of destructive collisions and the action of radiation
pressure. Assuming that their distribution  is the same as  an  underlying 
parent population of km-sized bodies, the apparent matter distribution  
within the disk
can be derived from numerical simulations of collisionless
particles. Such simulations  are able to reproduce the observed warp in the
inner part of the 
disk  as a result of the effect of gravitational  perturbations   due to a planet in an
orbit inclined at 3-5 degrees to the outer disk midplane. Such a planet should be located between 1 and
20 AU, with a corresponding mass respectively between $10^{-2}$ and
$10^{-5} M_{\star}$.

 
\section*{Acknowledgments}

This work was supported in part by EU grant ERB-CBRX-CT93-0329,
We are grateful to Alain Lecavelier and Alfred Vidal-Madjar for fruitful
discussions on the \bp ~disk and 
to N. Hubin and the ESO staff in La Silla for their help in the preparation
of and during the ADONIS observations.
%
%
%

\end{document}